\documentclass[usegraphicx,usenatbib]{mn2e}

\usepackage{fixltx2e}
\usepackage{mathptmx}

\newcommand{\apj}{ApJ}
\newcommand{\mnras}{MNRAS}
\newcommand{\nat}{Nat}

\newcommand{\aaps}{A\&AS}                 % "Astron. Astrophys. Suppl. Ser."
\newcommand{\aj}{AJ}                      % "Astron. J."
\newcommand{\pasp}{PASP}                  % "Publ. Astron. Soc. Pac."
\newcommand{\apjl}{ApJ}                   % letter at ApJ

\begin{document}

\title[CC of LSS]{Cross-Correlation of Diffuse Synchrotron and Large-Scale 
Structures}

\author[Brown et al]{Shea Brown$^{1,2,3}$, Damon Farnsworth$^{1}$, 
Lawrence Rudnick$^{1}$ \\
$^1$Department of Astronomy, University of Minnesota, Minneapolis, MN  
55455 \\ 
$^2$CSIRO, Australia Telescope National Facility, P.O. Box 76, Epping, NSW 
1710, Australia \\
$^3$Bolton Fellow, CSIRO, ATNF}
\maketitle

\begin{abstract} We explore for the first time the method of 
cross-correlation of radio synchrotron emission and tracers of large-scale 
structure in order to detect the diffuse IGM/WHIM. We performed a 
cross-correlation of a 34$^{\circ}\times34^{\circ}$ area of 2MASS galaxies 
for two redshift slices (0.03 $<$ z $<$ 0.04 and 0.06 $<$ z $<$ 0.07) with 
the corresponding region of the 1.4 GHz Bonn survey. For this analysis, we 
assumed that the synchrotron surface brightness is linearly proportional 
to surface density of galaxies.  We also sampled the cross-correlation 
function using 24 distant fields of the same size from the Bonn survey, to 
better assess the noise properties. Though we obtained a null result, we 
found that by adding a signal weighted by the 2MASS image with a filament 
(peak) surface brightness of 1 (7)~mK and 7 (49)~mK would produce a 
3$\sigma$ positive correlation for the 0.03 $<$ z $<$ 0.04 and 0.06 $<$ z 
$<$ 0.07 redshift slices respectively. . These detection thresholds 
correspond to minimum energy magnetic fields as low as 0.2~$\mu$G, close 
to some theoretical expectations for filament field values.  This injected 
signal is also below the rms noise of the Bonn survey, and demonstrates 
the power of this technique and its utility for upcoming sensitive 
continuum surveys such as GALFACTS at Arecibo and those planned with the 
Murchison Widefield Array (MWA).\end{abstract}


\section{Introduction} \vspace{-0.1in} In the current cosmological 
paradigm, $\approx$95\% of the mass/energy density of the universe is 
composed of dark energy and dark matter, both of which have yet to be 
directly detected. The remaining 5\% are ordinary baryons, which at 
redshifts of z $>$ 2 are fully accounted for based on Ly$\alpha$ forest 
observations of the photoionised intergalactic medium (IGM) and ordinary 
galaxies (e.g. Rauche et al. 1997; Weinberg et al. 1997; Schaye 2001). In 
the current epoch though, roughly half of these baryons are missing. 
Simulations suggest that the collapsing diffuse IGM was shock-heated and 
now resides in filaments as T$\sim$10$^{5}$ - 10$^{7}$~K WHIM, where it is 
practically invisible at most wavelengths (Cen \& Ostriker 1999, 2006; 
Dav{\'e} et al. 2001). A few tentative absorption detections of the 
coolest WHIM components using OVII and OVIII have been reported, but they 
give no information on the spatial distribution. However, shocks from 
infall into and along the filamentary structures between clusters are now 
widely expected to generate relativistic plasmas which track the 
distribution of the WHIM (Keshet et al. 2004; Pfrommer et al. 2006; Ryu et 
al. 2008; Skillman et al. 2008); indeed, merger/accretion shocks are seen 
as polarized radio sources (the ``peripheral relics") at the edges of the 
dense X-ray gas in clusters.  This radio emission also has the potential 
for probing into the lower density regions further from cluster cores. 
When such features are detected, they can be used to set limits on the
 pressure of the (invisible) thermal gas, delineate shock structures, and 
illuminate large scale magnetic fields. The next generation of low 
frequency radio telescopes (e.g., LOFAR, MWA, LWA, and eventually SKA), as 
well as current deep continuum surveys (e.g., 
GALFACTS\footnote{http://www.ucalgary.ca/ras/GALFACTS/} at Arecibo), have 
the potential to map this steep-spectrum, ``cosmic-web" of synchrotron 
emission (Battaglia et al. 2009).

However, there are two major problems that will challenge future radio 
surveys in their efforts to detect this emission. 1) Disentangling the 
extragalactic signal from diffuse Galactic emission that has higher 
surface brightness and similar angular scale; 2) Determining the redshift 
of the diffuse extragalactic signal (e.g., Brown \& Rudnick 2009).

One possible solution is to cross-correlate large field-of-view 
synchrotron maps with optical/IR tracers of large-scale structure (LSS; 
Keshet et al. 2004). The cross-correlation function (CCF) has long been 
used as means of detecting faint emission, often below the image sensitivity, 
in CMB maps such as the SZ and late ISW effects (e.g., Boughn \& 
Crittenden 2004). Cross-correlation can detect signals that are buried 
deep in the noise when the noise is not spatially correlated with the 
tracer image. In this case, the Galactic synchrotron ``noise" is not 
expected to be spatially correlated with the large-scale distribution of 
galaxies, while the synchrotron emitting relativistic plasma associated 
with the WHIM is correlated (Battaglia et al. 2009). 
Attempts to detect the WHIM with 
cross-correlation using the SZ effect has been tried with WMAP and 2MASS 
data (e.g., Hernandez-Monteagudo et al. 2004; Hansen et al. 2005; Cao et 
al. 2006), without success. In this paper we outline the first attempt to 
detect a correlation between synchrotron radio emission and large-scale 
structure, and demonstrate the power of cross-correlation by detecting 
simulated emission added into the synchrotron images. In this paper, we 
assume $H_{o}=70$, $\Omega_{\Lambda}=0.7$, $\Omega_{M}=0.3$.

\vspace{-0.3in} \section{Cross-Correlation}

\subsection{Images} Synchrotron emission in filamentary large-scale 
structure (LSS) is predicted to be on the order of 1-2~Mpc across, similar 
to some observed radio relics around rich X-ray clusters (see Giovannini 
\& Feretti 2004 for a review). For nearby redshifts, this corresponds to 
angular scales of (20$\arcmin$-40$\arcmin$)$\frac{0.04}{z}$. Therefore 
we used the Stockert 25m, 1.4~GHz Bonn survey \citep{reic82, reic86} with 
its 36$\arcmin$ resolution as a synchrotron tracer. This survey has an 
effective sensitivity of 50 mK (3$\times$rms noise); since it is a 
low-resolution, single dish survey, it is more sensitive to large-scale 
diffuse emission, and the relative contribution from point-sources is 
reduced significantly. We chose a target field, 
34$^{\circ}\times34^{\circ}$ centered on 16h00m00s +35d00m00s, where the 
diffuse Galactic emission was smooth and relatively well behaved. Figure 1 
shows the 1.4 GHz radio map, with 600 pixels of 3.4$^{\prime}$ each. Most 
of the signal in the image is diffuse Galactic emission, and the rest is 
dominated by systematic striping and unresolved radio galaxies.

\begin{figure*}
\includegraphics[width=0.6\textwidth]{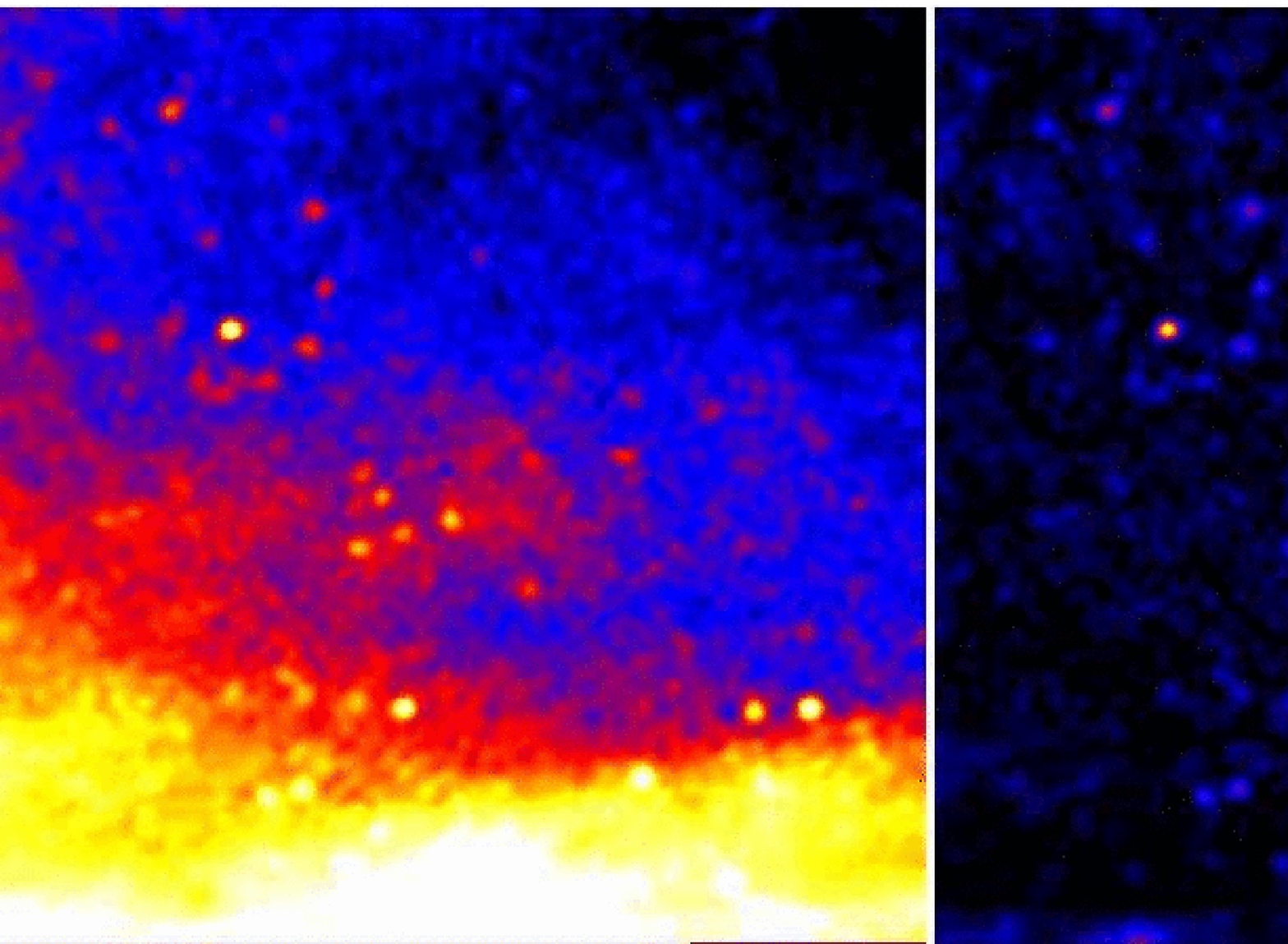}
\includegraphics[width=0.6\textwidth]{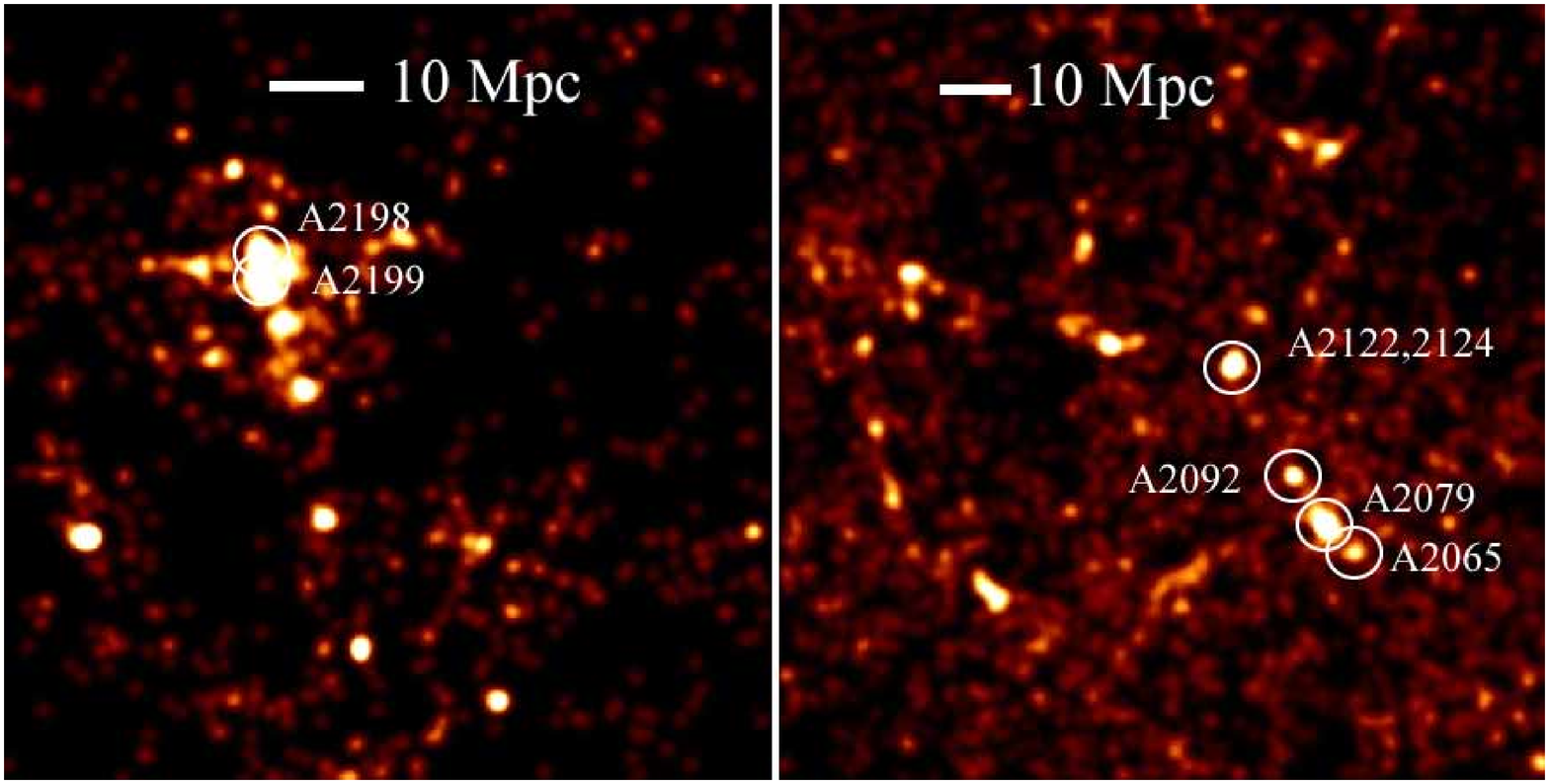}  

\caption{Center coordinates for each image are 16h00m00s +35d00m00s.  
Top-Left: Original 34$^{\circ}\times34^{\circ}$ 1.4 GHz, intensity Bonn 
image; Top-Right: Same field with Galactic gradients removed. The brighter 
point sources are 400-600~mK above the 3-3.5 K background in the 
unfiltered image. Bottom-Left: The distribution of galaxies from 2MASS 
with 0.03 $<$ z $<$ 0.04, convolved to 36$^{\prime}$. Peak/mean=64; 
Bottom-Right: Same for 0.06 $<$ z $<$ 0.07. Peak/mean=30.}

\end{figure*}

We filtered out the diffuse Galactic synchrotron emission using the 
multi-scale filtering technique of Rudnick (2002). The choice of filtering 
size is a compromise between the need to remove as much as possible of the 
Galactic emission while preserving the filamentary component. For this 
initial experiment, we filtered out all emission larger than 
180$^{\prime}$, or 5 times the size of the Bonn beam. As long as the 
filaments are significantly narrower than 180$^{\prime}$, they will not be 
filtered out, independent of their length (see, e.g., results on jets in 
Rudnick, 2002). Further experimentation to optimize this will be important 
for future studies. Figure 1 shows the target field after filtering.

For a tracer of LSS in our target field, we used the 2MASS survey 
(Skrutskie et al. 2006) to create maps of the surface density of galaxies 
(number of galaxies/pixel) for 2 redshift slices (0.03 $<$ z $<$ 0.04 and 
0.06 $<$ z $<$ 0.07). We then convolved the 2MASS maps to 36$^{\prime}$, 
the resolution of the Bonn survey. Figure 1 shows the 2MASS images. The 
large concentration of galaxies in the 0.03 $<$ z $<$ 0.04 image 
corresponds to the two clusters Abell 2199 and 2197 at redshifts of 0.0301 
and 0.0308, respectively. The 0.06 $<$ z $<$ 0.07 image contains Abell 
clusters 2122, 2124, 2079, and 2065 at redshifts of 0.0661, 0.0656, 
0.0670, and 0.0726, respectively.

\subsection{Cross-Correlation} We computed the cross-correlation function 
(CCF) for the filtered and unfiltered radio images with both redshift 
slices. The CCF is defined by \[ CCF\left(xshift, yshift\right) = 
\frac{1}{n} \sum (R_{i,j}-\bar{R}) (G_{i,j}-\bar{G}), \] where $n$ is a 
normalization given by $\sqrt{\sum (R_{i,j}-\bar{R})^{2} \sum 
(G_{i,j}-\bar{G})^{2}}$, R is the radio map, G is the convolved 2MASS 
surface density of galaxies shifted in relation to R by $xshift$ and 
$yshift$ pixels, and the sum is over every pixel \{i,j\} in the 
overlapping region of the images. $\bar{R}$ and $\bar{G}$ are average 
values for the radio and galaxy surface-density maps respectively. Figure 
2 shows the CCF, with the shift in units of Mpc at the redshift of the 
2MASS images, for both redshift slices cross-correlated with the Bonn 
images. It is obvious that the large-scale gradient in the Galactic 
synchrotron emission is dominating the CCF in the unfiltered images. The 
gradient is due to the fact that the power in the 2MASS galaxy 
distributions are concentrated in a few small regions. This is especially 
evident at 0.03 $<$ z $<$ 0.04, where there is one dominant clump of galaxies 
which dominated the cross-correlation signal and causes the gradient when 
correlated with the Galactic synchrotron gradient. The CCF for the 
filtered images is not significantly peaked at zero shift, i.e., the 
filtered synchrotron and 2MASS surface densities are not significantly 
correlated.

When cross-correlating all-sky images, such as done with the CMB, the 
significance of a given CC value is assessed by using simulated CMB 
signals (e.g., McEwen et al. 2007).  In the initial experiments peformed 
here, we are using only a 34$^{\circ} \times$ 34$^{\circ}$ field, and can 
therefore use other fields from the Bonn survey as additional controls 
to better assess the noise properties.  This is important because
of  the systematic striping in the Bonn survey, and the fact that diffuse Galactic emission is not a simple Gaussian field like the CMB. This initial study 
sampled the cross-correlation function using fields from the Bonn survey 
at distances ($\theta$) ranging from 27$^{\circ}$ to 127$^{\circ}$ from 
our target field. We split the survey up into 34$^{\circ}\times34^{\circ}$ 
images with 600 pixels on a side and TAN projection (Greisen \& Calabretta 
2002; Calabretta \& Greisen 2002), and excised all the fields with strong 
Galactic plane emission. Fields with significant artifacts from gridding the survey 
maps together were also removed, leaving 24 total images. Assuming that 
none of the 24 maps should be correlated with the surface density of 
galaxies in a different, well separated region of the sky, the 
distribution of these CC values (histograms are shown in Fig. 3) 
represents the noise in this procedure. The CCF($\theta=0$) value for the 
target field is marked with a red line; in all cases the 2MASS-Bonn 
correlation for the target field is again not significant. Note that the 
scatter is dramatically lower using filtered images.

One can think about this procedure as constructing a sparsely sampled 
CCF($\theta$) over the whole sky. Figure 4 shows plots of the CCF as 
sampled by the 24 distant fields as a function of angular distance from 
the target field (which here is CCF($\theta=0$)). These again show that 
the target CC value is not significantly above the values for the 
well-separated control fields.

\begin{figure*}
\includegraphics[width=4.2cm]{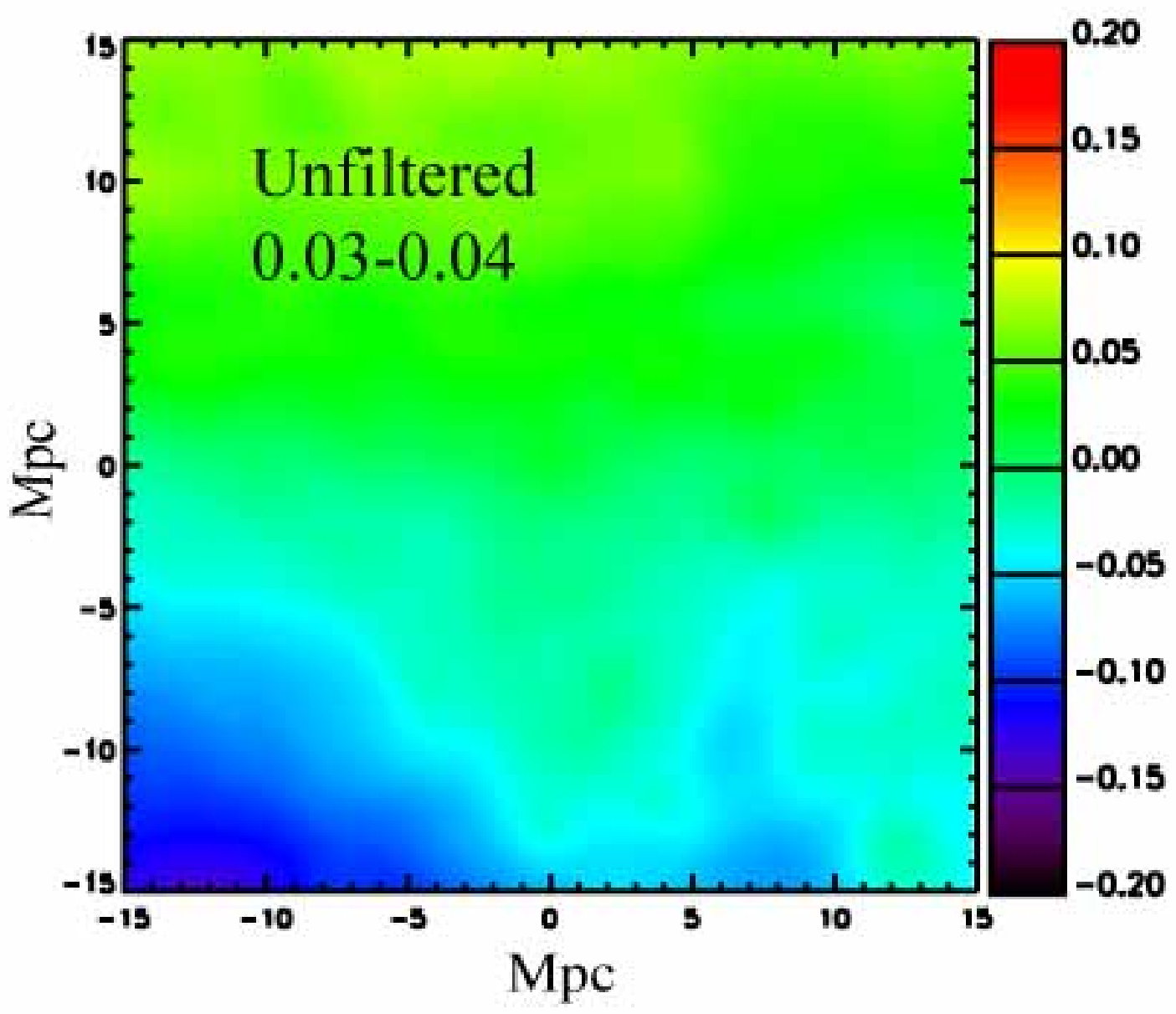}
\includegraphics[width=4.2cm]{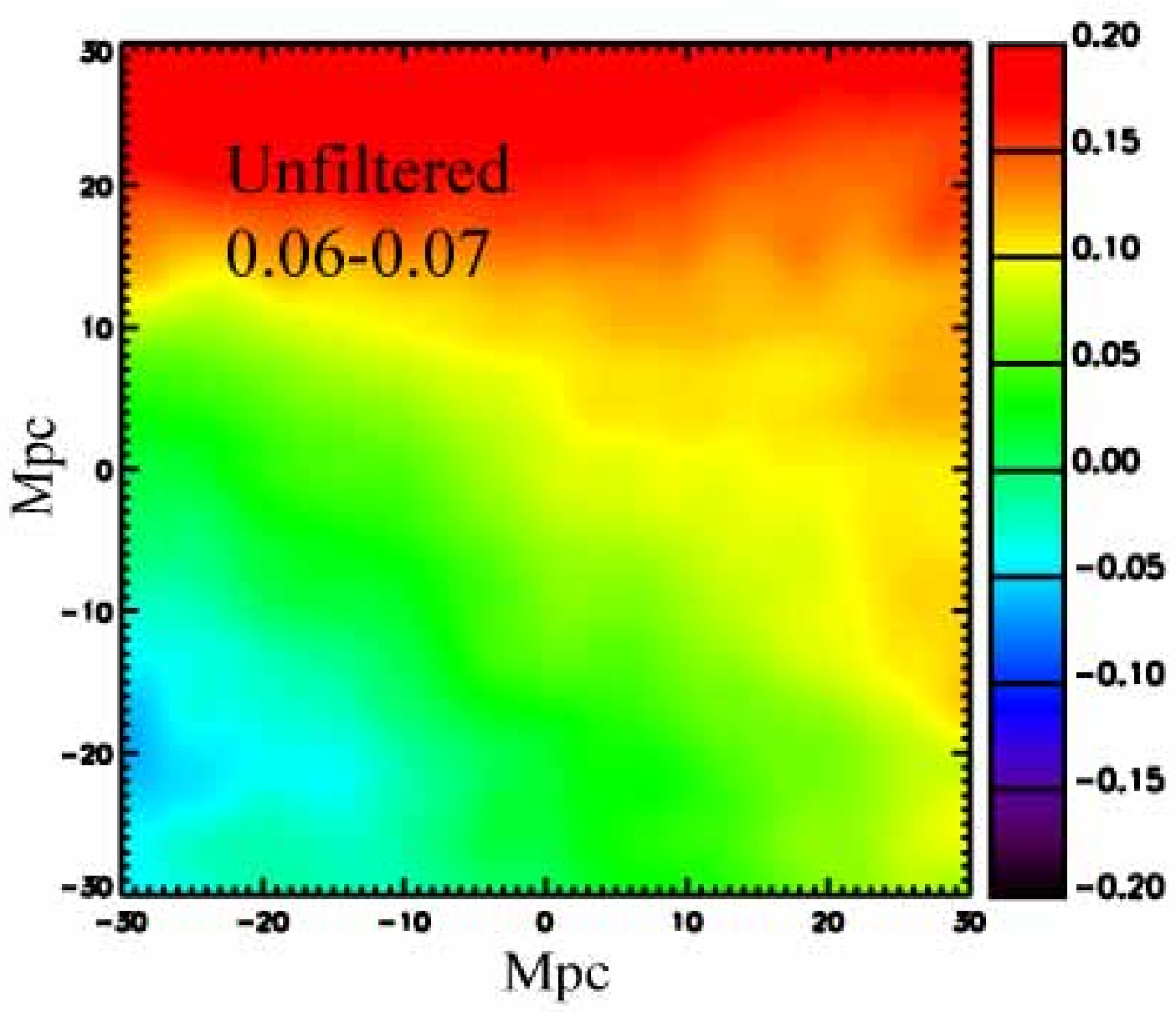}
\includegraphics[width=4.2cm]{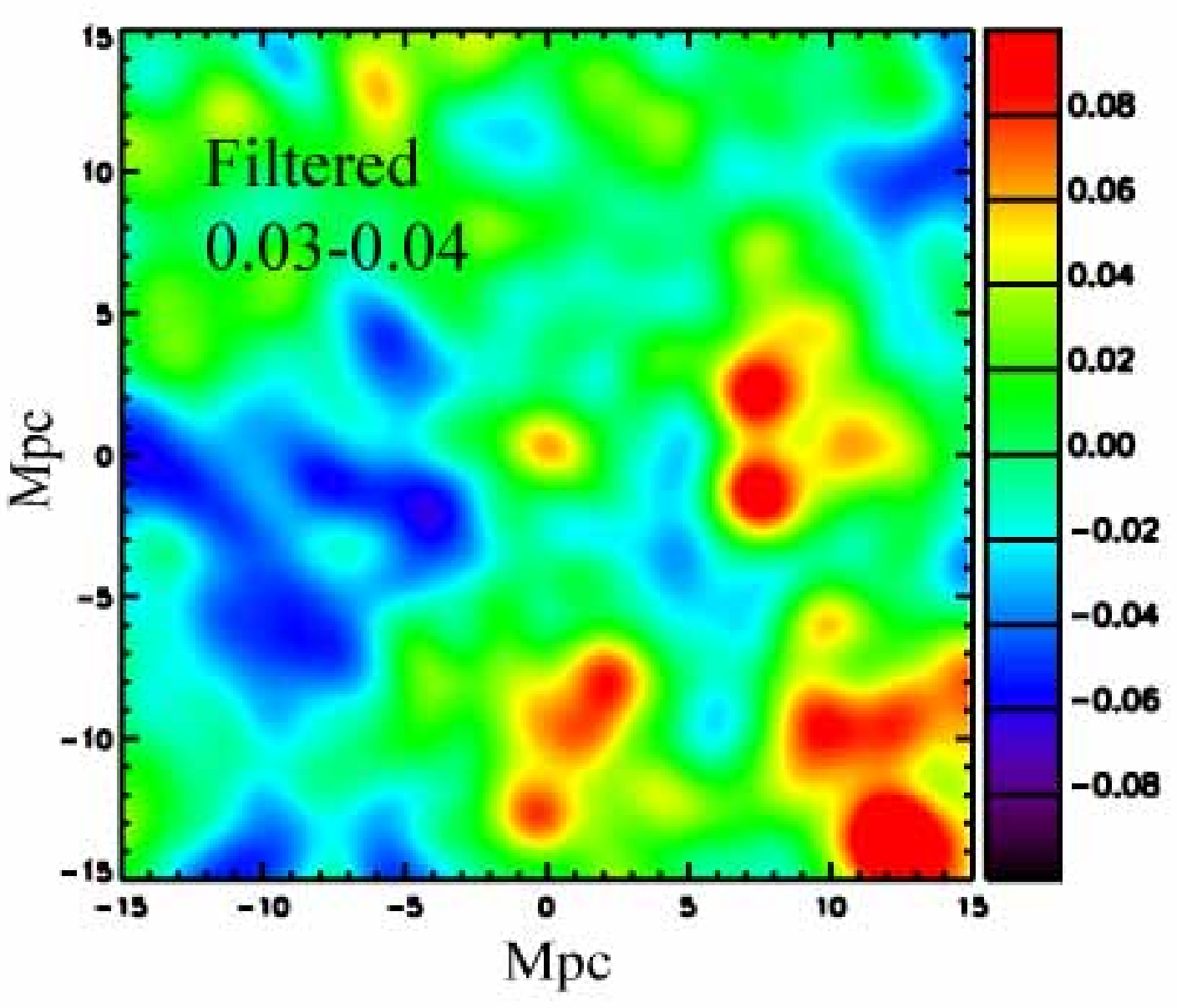}
\includegraphics[width=4.2cm]{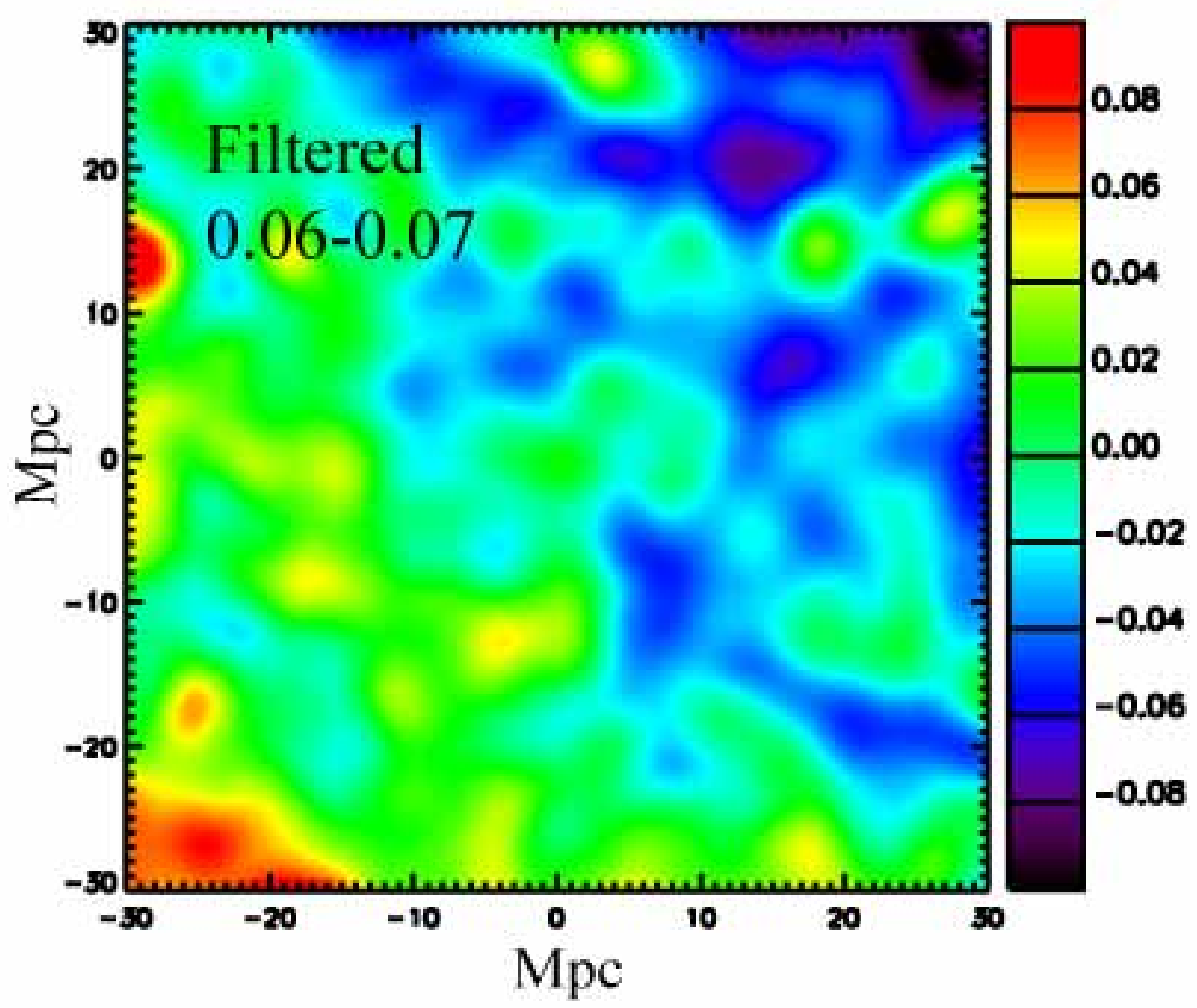}

\caption{CCF for the unfiltered and filtered 1.4 GHz total intensity Bonn 
images and the 0.03 $<$ z $<$ 0.04/ 0.06 $<$ z $<$ 0.07 2MASS surface 
density of galaxies. The angular scale for each image is 11.3$^{\circ}$ on 
a side (total shift of 201 pixels), which is $\sim$19 independent beams 
across.} 

\end{figure*}

\begin{figure}
\begin{center}
\includegraphics[width=7cm]{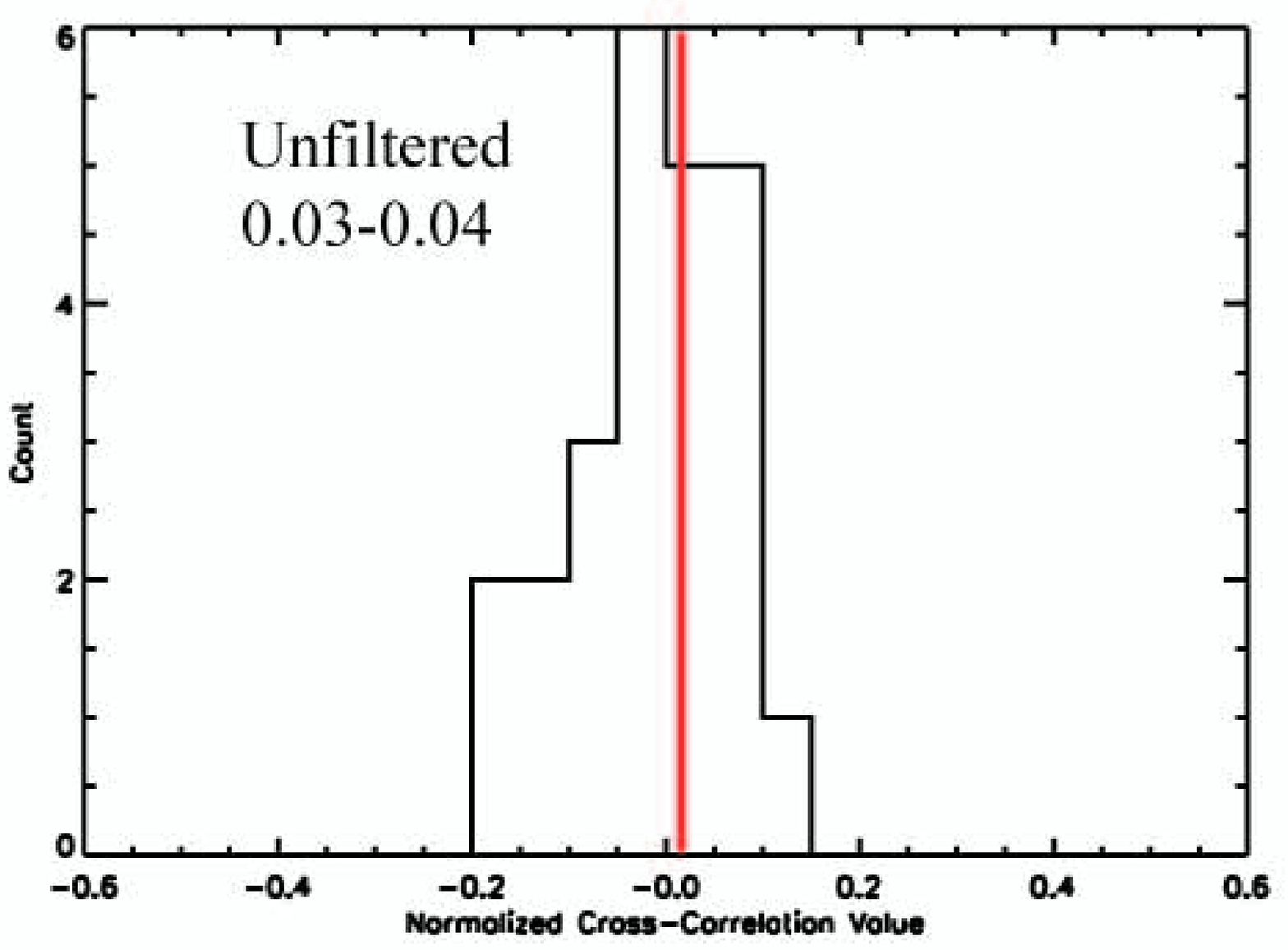}
\includegraphics[width=7cm]{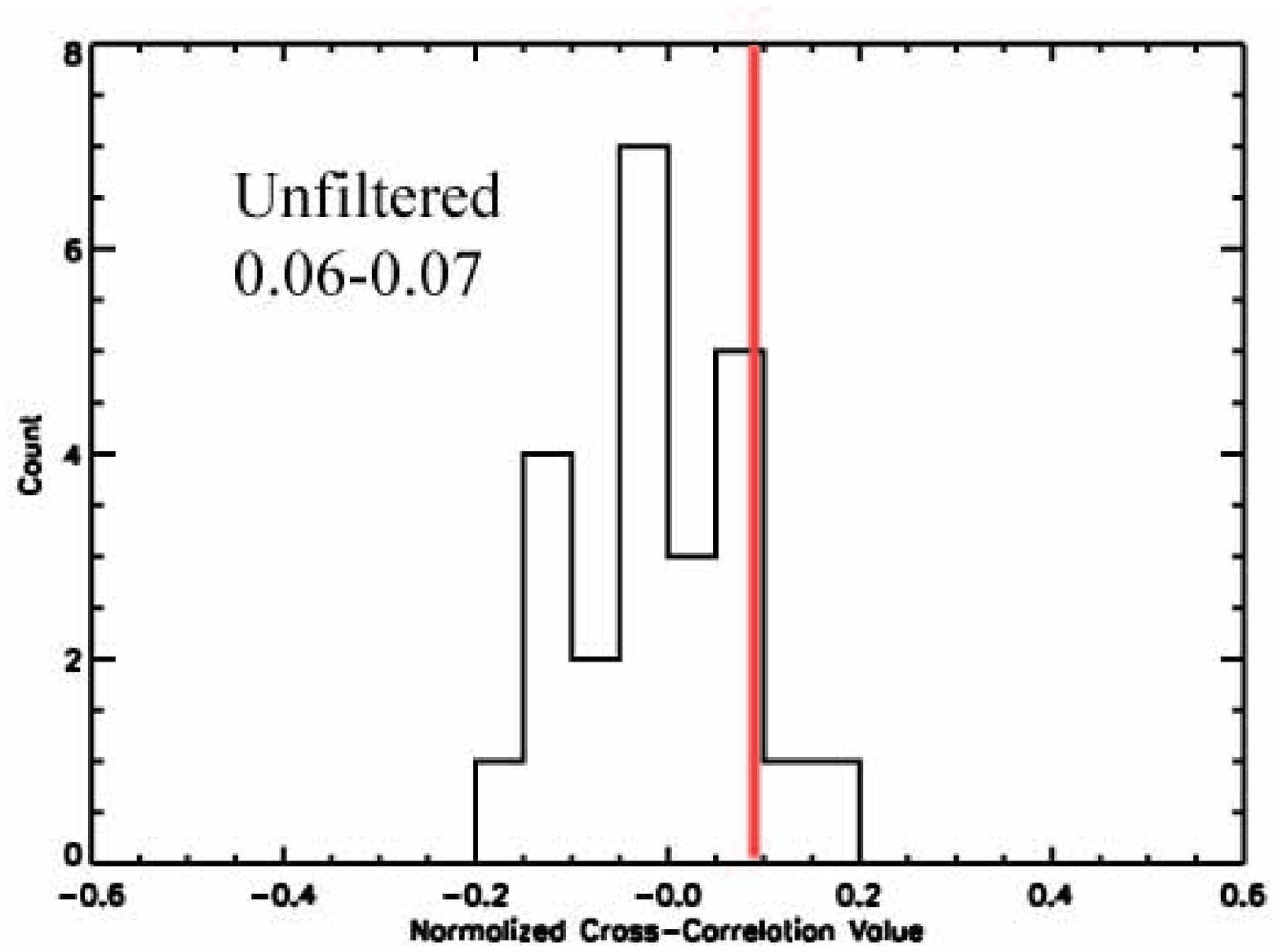}
\includegraphics[width=7cm]{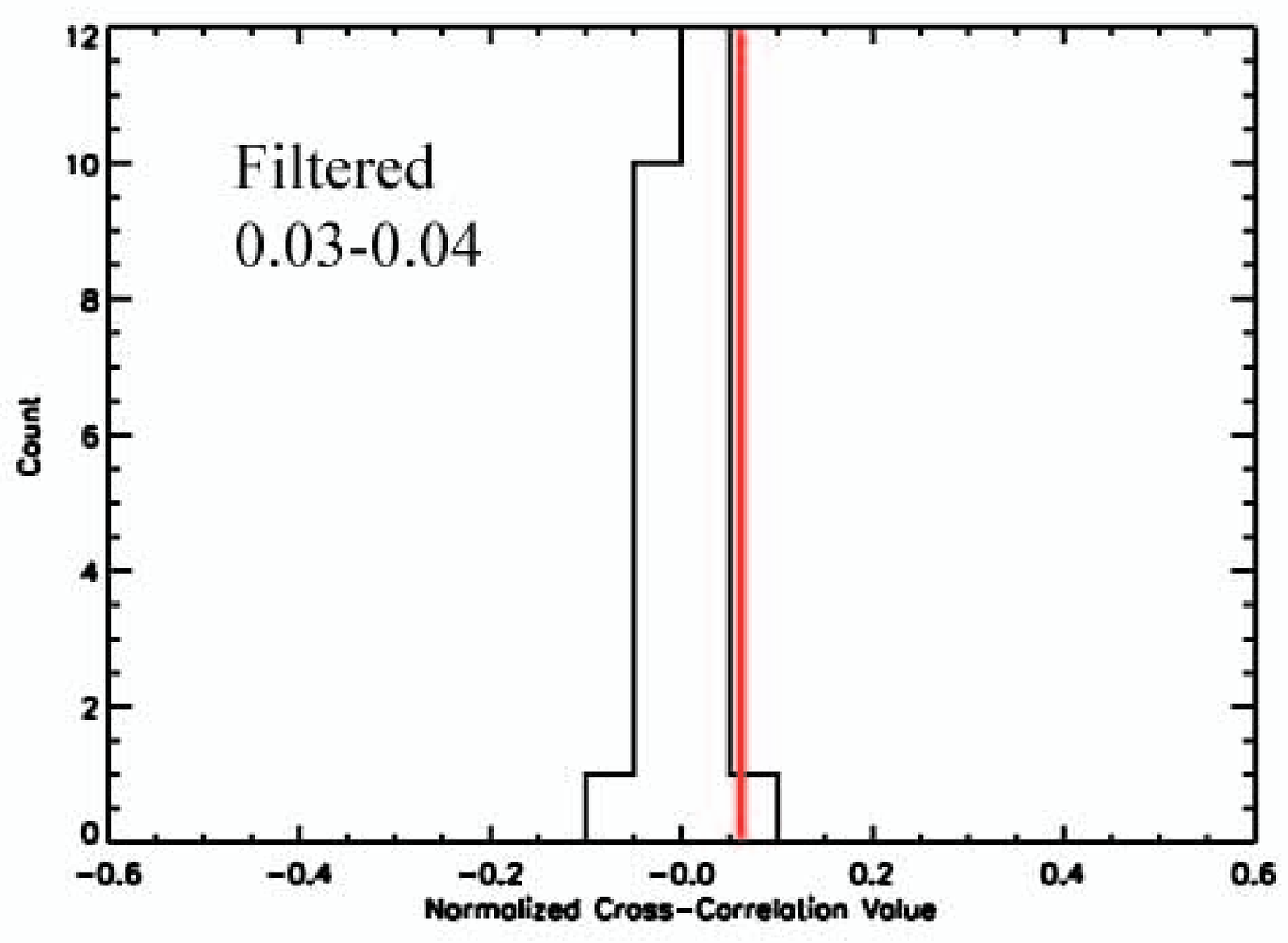}
\includegraphics[width=7cm]{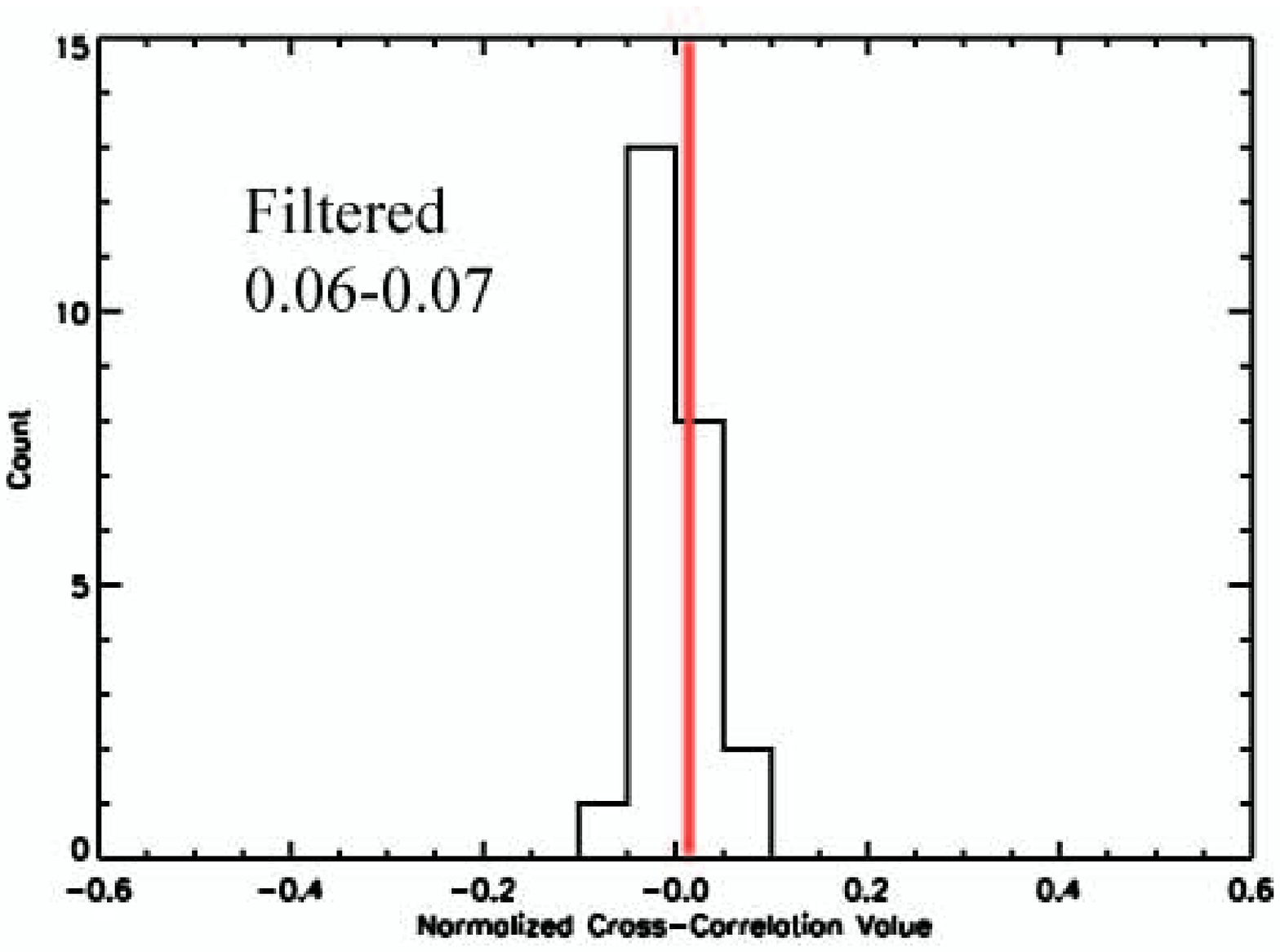}

\caption{Distribution of cross-correlation values ($\theta$=0) for the 
Bonn images with the 2MASS galaxy distribution images. The solid lines 
indicate the CC value of the target field centered on 16H35D.}

\end{center}
\end{figure}

\begin{figure}
\begin{center}
\includegraphics[width=8cm]{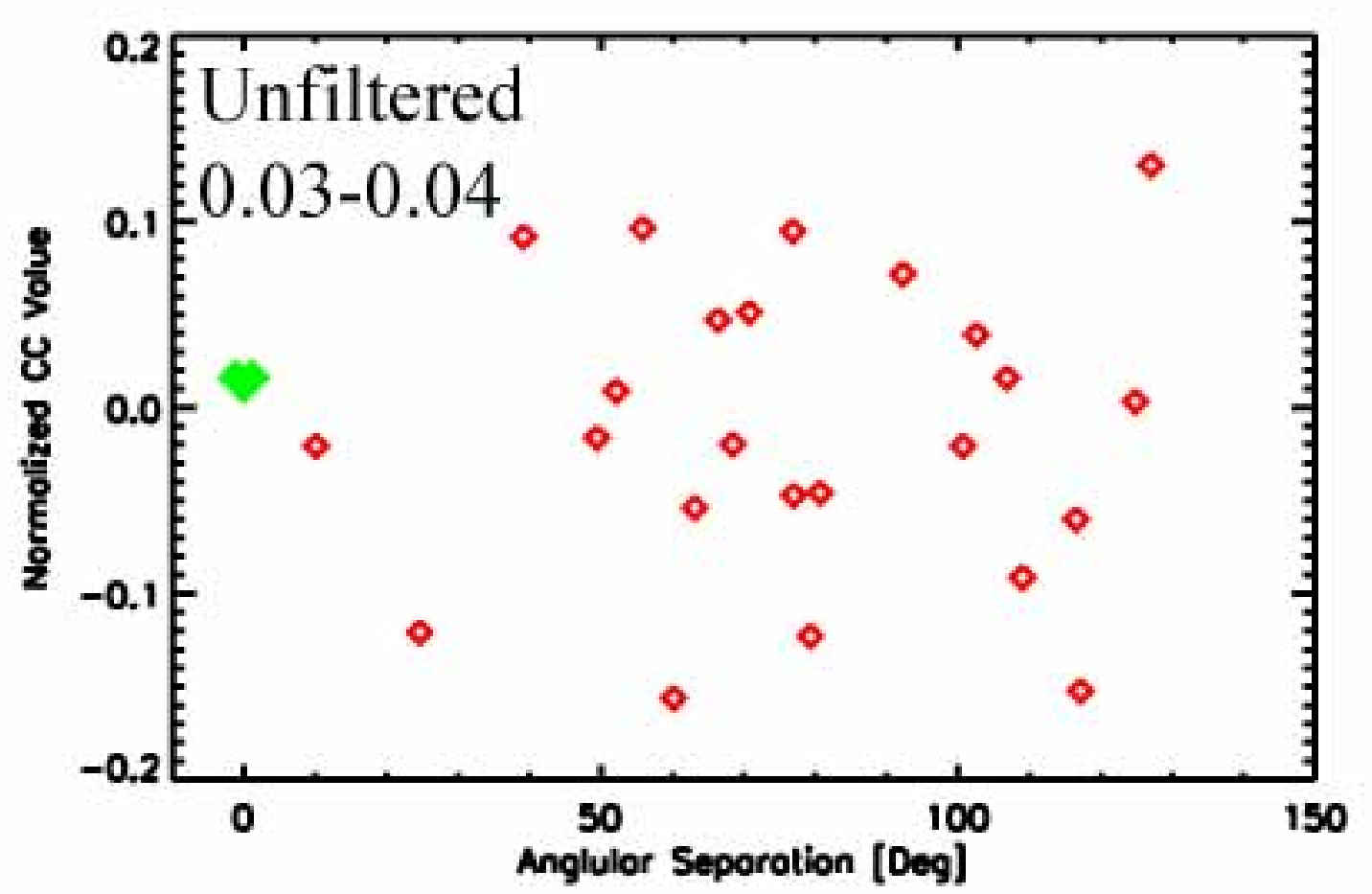}
\includegraphics[width=8cm]{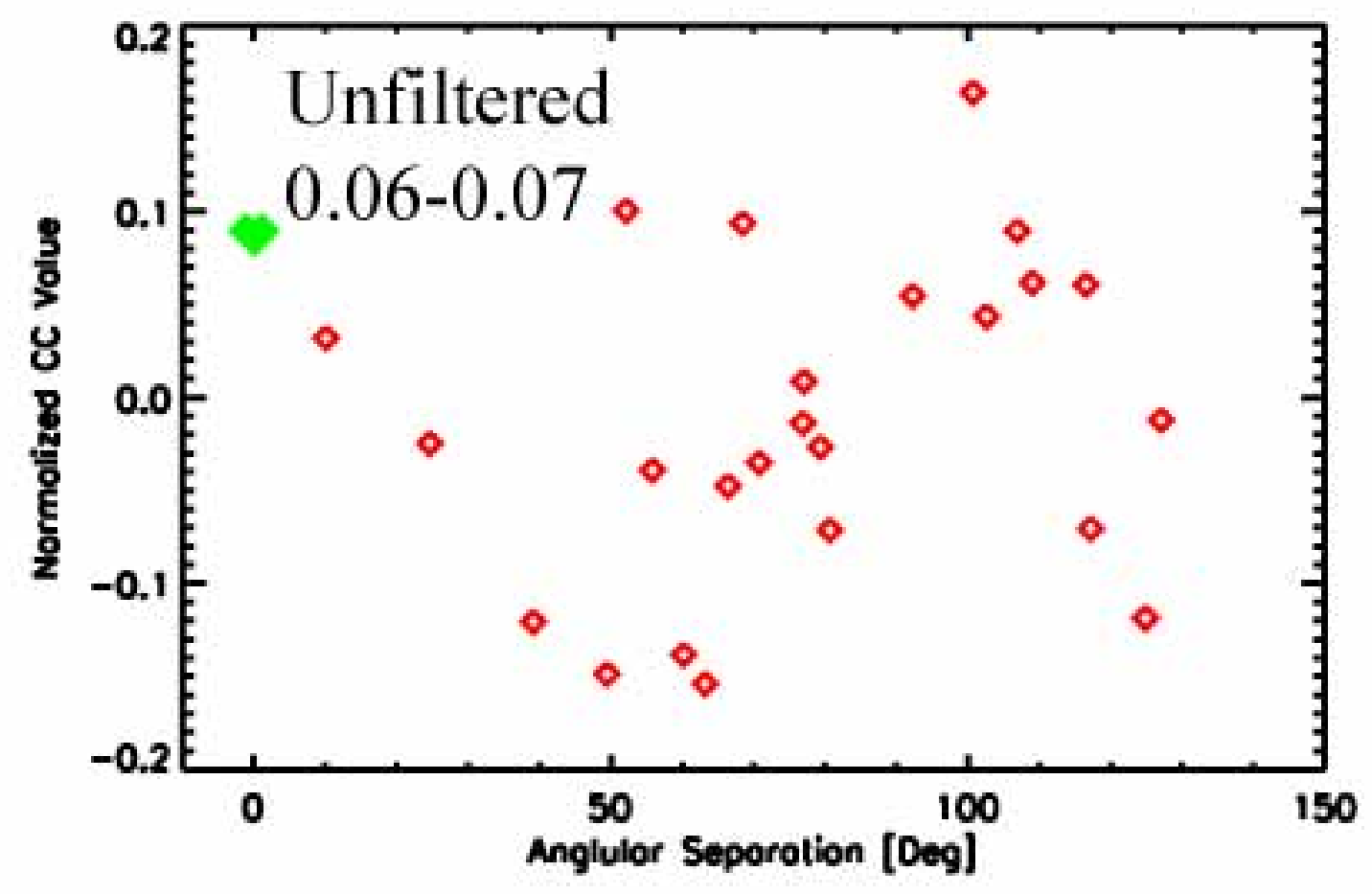}
\includegraphics[width=8cm]{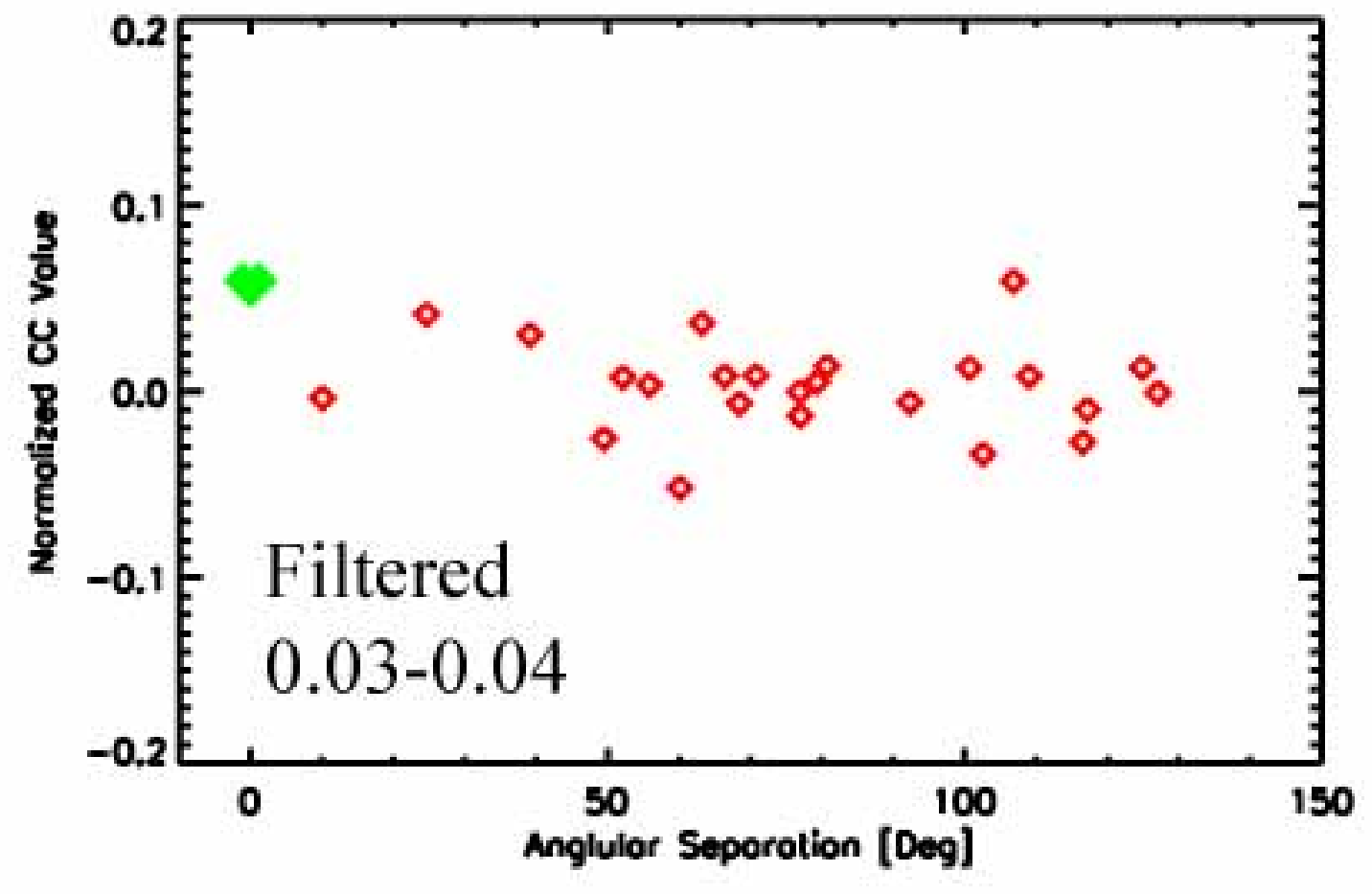}
\includegraphics[width=8cm]{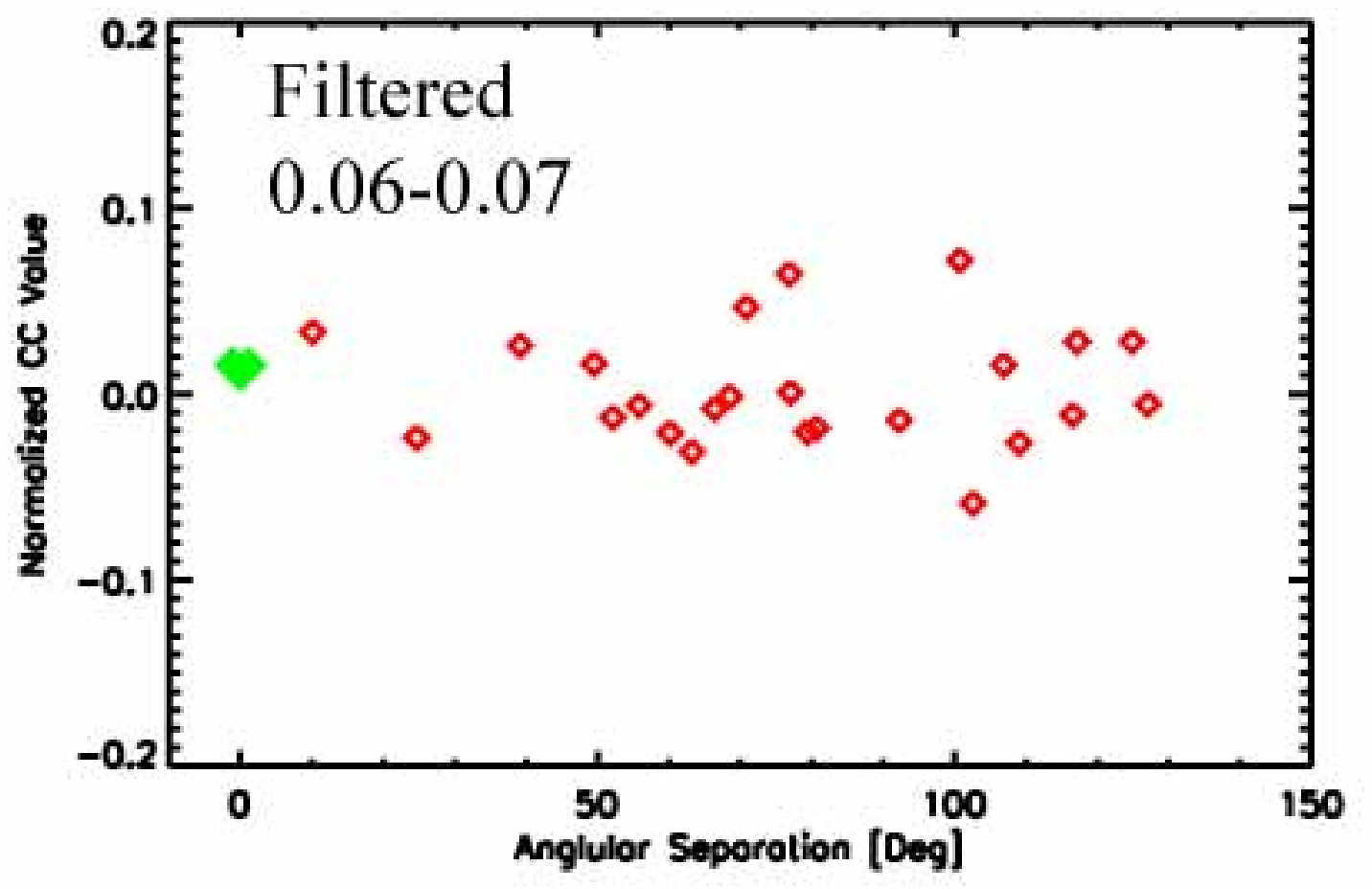}

\caption{Sparsely sampled CC functions. Open diamonds are the 2MASS images 
correlated with the 24 Bonn fields. The large solid diamonds are the 
target fields CC value.}

\end{center}
\end{figure}

\subsection{Detection Thresholds} In order to assess our sensitivity to 
diffuse synchrotron signals associated with large-scale structure, we next 
injected a simulated filamentary, LSS radio signal into the Bonn images 
before cross-correlation. As a simple model, we assumed the synchrotron 
LSS signal is linearly proportional to the number-density of galaxies. Our 
model assumes that each galaxy in the redshift bin emits a certain number 
of mJy at 1.4 GHz (the same for each galaxy), which is then convolved by 
the 36$^{\prime}$ Bonn beam and added incrementally into the original Bonn 
radio map. The injected signal is a clumpy distribution of 
surface-brightness that traces clumps of galaxies, and mimics the bottom 
two panels in Fig. 1 exactly except for an overall multiplicative factor. 
The detection threshold is defined as the number of mJy/galaxy needed to 
get a cross-correlation signal at $\theta$=0 that is 3 times the rms 
distribution of the sparsely sampled CC function (Figures 3 \& 4). We did 
this separately for both the filtered and unfiltered Bonn images and both 
redshift ranges. The peak and typical filamentary structure values of the 
injected signal in mK are also listed in Table 1. In the case of the 
filtered Bonn image correlated with each redshift, the peak signal needed 
to cause a 3$\sigma$ detection is \emph{below} the 3$\times$rms 
sensitivity level of the Bonn survey, because many different independent 
beams are included in the CC.

We also looked briefly at the issue of isolating the CC contribution of 
the dense cluster-like concentrations in 2MASS from the contribution from 
more diffuse filamentary structures. To explore this, we repeated the 
signal injection experiment with the z=0.03-0.04 image and the filtered 
Bonn image, except that this time we removed the bright clumps by only 
adding in signals that were below 5$\sigma_{rms}$ of the 2MASS image. In 
this case the detection threshold rises significantly, to peak and 
filament values of 19 mK and 11 mK, respectively. We note that these 
values are still well below the sensitivity of the Bonn survey for any 
given location.

\begin{table*} 
\begin{minipage}{13cm}
\centering
\caption{3$\sigma$ Cross-Correlation Detection Thresholds}
\begin{tabular}{@{}lcccccc}
\hline
 & Unfiltered 0.03$<$z$<$0.04 & Unfiltered 0.06$<$z$<$0.07 
 & Filtered 0.03$<$z$<$0.04$^{a}$ & Filtered 0.06$<$z$<$0.07 \\
& (mK) & (mK) & (mK) & (mK)\\
\hline
Peak & 694 & 570 & 7[19] & 49 \\
Filament & 100 & 80 & 1[11] & 7 \\  
\hline
mJy/Gal$^{b}$ & 205 & 135 & 2 & 12 \\
\hline
Filament B$_{eq}$ ($\mu$G) & 0.74f$^{-2/7}$ & 0.70f$^{-2/7}$ & 0.20f$^{-2/7}$ & 0.35f$^{-2/7}$ \\
\hline
\end{tabular}
a: Bracketed values are thresholds when including signals from filaments only (see 
text).\\
b: The flux required per galaxy to reach the detection threshold. \\
c: f is the volume filling factor of the synchrotron emission
\end{minipage}
\end{table*}

\section{Physical Expectations/Interpretations} How do these simulated 
radio signals compare to theoretical expectations for filamentary 
synchrotron emission? Resent cosmological simulations (Ryu et al 2008) 
demonstrated that strong turbulence driven by structure shocks associated 
with the WHIM can generate current epoch (volume averaged) filament 
magnetic fields of the order of 0.01~$\mu$G. However, the synchrotron 
emissivity scales as B$^{2}$, and under the assumption that the total 
number of relativistic particles will scale with local mass density, the 
mass weighted rms of B is appropriate. This yields a considerably higher 
characteristic value of 0.2$\mu$G in filaments (Ryu, private 
communication, 2008). In order to convert this magnetic field estimate to 
a characteristic synchrotron brightness, we need to make some assumption 
about the cosmic ray density. We adopt the condition where the energy 
density in relativistic particles (protons and electrons, equally) is the 
same as the magnetic field energy density, (equipartition). This results 
in a synchrotron brightness at 1.4 GHz of 1~mK in filaments. This 
predicted value is much weaker than the $\le$50mK background brightness 
temperature predicted by \cite{kesh04}. We note that there is considerable 
uncertainty regarding the actual magnetic strength in filaments.  Donnert 
et al. (2008), e.g., find only $\sim$0.005~$\mu$G.

Table 1 lists the equipartition magnetic fields corresponding to our 
detection thresholds.  Our lowest thresholds approach those expected in 
the Ryu et al. (2008) simulations, when the appropriate mass-weighted
rms averaging is done.  With more sensitive future 1.4~GHz 
surveys such as GALFACTS, we will be able to robustly test those 
expectations, although it is not yet clear whether we can reach an
average field strength of 0.005 $\mu$G (Donnert et al. 2008).

 To put our CC brightness limits in context, we note that even bright 
radio halos such as in the Coma cluster (with $\sim$20$^{\prime}$ FWHM, 
e.g., Kim et al. 1989) are very difficult to detect individually. The Coma 
halo in the Bonn survey is a barely detected 150~mK bump on the 
$\sim$3.3~K CMB + Galactic local background, mostly due to the fact that 
the brightness is diluted by the 36$^{\prime}$ beam. Typical currently 
known halos have brightnesses of $\approx$700~mK, brighter than the 
\emph{intrinsic} 350-400~mK brightness of the Coma cluster halo.  As 
discussed in Rudnick \& Lemmerman (2009), visual searches for halos are 
biased towards higher brightnesses at higher redshifts.

The fact that the detection thresholds (Table 1) are below the predicted 
values of Ryu et al. (2008) and Keshet et al. (2004) do not rule out these 
models. Important effects such as the filling factor of the synchrotron 
emission and the linear nature of the LSS formation shocks are not 
captured by our simple model. However, analysis of LSS formation 
simulations that include synchrotron emission can be used in the future to 
design the optimum CC procedure for detecting this emission, and would 
guide the way toward the correct interpretation of future 
cross-correlation results. The ultimate goal is to provide limits on 
magnetic field strengths and pressure of the relativistic plasma in 
filaments, and therefore test origin theories for cosmic magnetism and 
probe the missing baryons in the WHIM.

\section{Summary} We have explored for the first time the method of 
cross-correlation of diffuse radio emission and tracers of LSS in order to 
detect the synchrotron radiation associated with the diffuse baryons in 
the WHIM. Filtering out the large-scale Galactic emission is critical for 
reducing the CC noise. We demonstrated that, if the filamentary radio 
signal is traced perfectly by the distribution of galaxies, emission 
significantly below the sensitivity of the Bonn survey can be 
statistically detected. Our detection thresholds are already at a level 
expected in some models, and dramatic improvements using new surveys will 
be possible in the future.

Further refinements to the CC technique, including subtraction of point 
sources which also trace the LSS, will also improve its use as a detection 
tool for the diffuse baryons in filaments. The assumption of perfect 
correspondence between the distribution of galaxies and IGM radio emission 
is clearly an oversimplification, and further experimentation with 
numerical simulations is needed.

Partial support for this work at the University of Minnesota comes from 
the U.S. National Science Foundation grant AST~0607674. We acknowledge the 
use of NASA's SkyView facility located at NASA Goddard Space Flight 
Center. SB also acknowledges support from a Doctoral Dissertation 
Fellowship from the Graduate School of the University of Minnesota.

\end{document}